# Space-time non-separable dynamics of ultrashort vortex pulse with power exponential spectrum

**Shakti Singh[1] and Akhilesh Kumar Mishra[1, 2, *]**

[1]*Department of Physics, Indian Institute of Technology Roorkee, Roorkee-247667, Uttarakhand, India*
[2]*Centre for Photonics and Quantum Communication Technology, Indian Institute of Technology Roorkee,*
**Corresponding author akhilesh.mishra@ph.iitr.ac.in*

**Abstract:** Ultrashort optical pulses with orbital angular momentum (OAM) owing to their applications in classical as well as quantum domains attract a lot of research attention. The evolution of their spatio-temporal dynamics is of particular interest. In the present work, the spatio-temporal evolution of ultrashort Laguerre Gauss (LG) vortex with power exponential spectrum, also called power exponential (PE) pulse is numerically explored, in dispersive and nonlinear media. The evolution of space-time non-separable ultrashort vortex pulse is specifically addressed. The study reveals the asymmetric spatio-temporal evolution and splitting in the spatio-temporal plane. Compression at the trailing edge of the pulse is observed with increase in the strength of the nonlinearity. The study further reveals that the temporal, spectral, and chirp evolution patterns of the pulse vary uniquely across different spatial regions and these variations depend upon the strength of nonlinearity and propagation distance. In weak nonlinear regime, a redshift in the spectrum is observed as one moves radially outward from the bright caustic of the ultrashort LG vortex, while a blueshifted spectrum emerges as one approaches the phase singularity. Interestingly, as the pulse furthers in the medium, nonlinearity induced spectral shift is witnessed and with stronger nonlinearity, a complex spectral evolution is observed.

## 1. Introduction:

Ever since the advent of ultrafast laser sources, due to their unprecedent properties and related applications the field of ultrafast optics has witnessed continuing research interest. The ultrashort pulses have found significant attention across various research areas such as metamaterials [1], communications [2], particle acceleration [3], laser machining [4], nonlinear and topological photonics [5] to name a few. In these mentioned applications, it is important to precisely control the shape of optical pulses not only in time but in space as well. Consequently, recent years have witnessed a growing interest in understanding the dynamics of optical pulses in both the dimensions.

Generally, optical pulses in factorized solution of the Maxwell's equations are treated as space time separable solution, where a pulse can be expressed as product of spatial function and temporal function. However, this assumption does not hold good in ultrafast regime and hence a non-separable space-time solution was proposed [6]. In ultrafast domain, optical pulses exhibit a correlation between spatial and temporal degrees of freedom, known as spatio-temporal coupling [7]. Ultrashort optical pulses, due to their broad spectral bandwidth, inherently exhibit spatio-temporal coupling. This arises from angular dispersion, which distorts pulses more as their bandwidth increases. The consequence of spatio-temporal coupling leads to emergence of phenomena such as iso-diffraction and iso-divergence [8]. In ultrashort pulse, the spatio-temporal coupling results phenomenal effects such as red shift in the spectrum, time dependent diffraction, space dependent pulse time delay etc [9–11].

Since the first realization that Laguerre Gauss (LG) beam with azimuthal integer $l$ carries an orbital angular momentum (OAM) of $l\hbar$ per photon, the exploration of light carrying OAM has surged in recent years [12–14]. Light with OAM exhibits undefined phase at its centre called phase singularity. Light with OAM finds applications in various field such as optical communication [15], micromanipulation [16], phase-contrast microscopy [17], and quantum information processing [18] among others.

In recent years, ultrashort pulses with OAM (also known as ultrashort vortices) have also been investigated theoretically as well as experimentally [19–21]. As stated earlier ultrashort pulses exhibit spatio-temporal coupling. While ultrashort vortices possess additional coupling between OAM degree of freedom and temporal degree of freedom due to which temporal shape strongly depends upon OAM [22,23]. The temporal-OAM coupling is a special case of spatio-temporal coupling. Researcher have been familiar with spatio-temporal coupling of the wave equation solutions for decades. Typically, these couplings are small in ultrashort pulses without OAM [24]. However, it can be amplified for specific applications artificially [7]. In ultrashort vortices, high intense spatial regions and low intense spatial regions exist and temporal-OAM coupling varies considerably in these spatial regions [22]. Recently, it has been demonstrated theoretically that under iso-diffracting condition, ultrashort vortices maintain a propagation invariant temporal shape [20]. Importantly, it has also been reported that ultrashort LG vortices and X-waves cannot carry arbitrary amount of OAM, rather there exists a lower bound to their temporal duration for given value of $l$ [20,23]. Therefore, optical vortices with certain available spectral bandwidth cannot carry arbitrary amount of OAM. M. A. Porras et. al. has reported that for ultrashort LG vortices and X waves, the spectrum blueshifts as one moves towards the singularity and redshifts on traversing radially outward from the bright caustic [22,25]. Ultrashort optical pulses with OAM have been extensively studied in free space as well as dielectric media in space time separable case. Space-time nonseparable ultrashort optical pulses in toroidal and supertoroidal form have also been investigated in detail [26-28]. However, to best of our knowledge space-time non-separable ultrashort optical pulse with OAM in dielectric media have not been investigated yet.

In this work, we numerically investigate the dynamics of ultrashort LG vortex with PE spectrum in dispersive and nonlinear media. For modelling the propagation of the ultrashort vortex pulse, we solve the three-dimensional nonlinear Schrödinger (NLSE) equation. In ultrashort domain, however, the conventional slowly varying envelope approximation (SVEA) fails. Hence, we employ the nonlinear envelope equation (NEE) to get the deeper insights into the vortex pulse dynamics [29]. In section 2 we discuss the theory of wave equation where we see that when the temporal duration of pulse becomes comparable to the time period corresponding to the carrier frequency of the pulse, Maxwell's equations give non-separable spatio-temporal solutions. Section 3 discusses the numerical results of this paper. In this section we explore the spatio-temporal, spectral, and frequency chirp evolutions of ultrashort LG vortex across different levels of nonlinearity and propagation distances at different transverse regions of the vortex. Section 4 concludes the work.

## 2. Theory and Model:

Classically, light propagation through different kind of media is governed by Maxwell's equations. The mathematical expression of the wave equation in free space is derived from Maxwell's equations in scalar form is given by,

$$\nabla^2 E - \frac{1}{c^2}\frac{\partial^2 E}{\partial t^2} = 0, \tag{1}$$

where $E$ is the electric filed, $c$ is the speed of light and $\nabla^2$ is Laplacian operator. Let's consider the electric field as $E = \psi(x, y, z, t)e^{i[\omega_0(t-z/c)]}$, where $\omega_0 = 2\pi/T_0$ is carrier frequency and $T_0$ is the corresponding time period. Substituting this expression of field in equation (1) under scalar approximation, we get

$$\nabla^2\psi - 2i\frac{\omega_0}{c}\frac{\partial\psi}{\partial z} - 2i\frac{\omega_0}{c^2}\frac{\partial\psi}{\partial t} - \frac{1}{c^2}\frac{\partial^2\psi}{\partial t^2} = 0. \tag{2}$$

Let's consider the transformation $\tau = t - z/c$ and $z' = z$,

$$\nabla^2_{x,y}\psi - 2ik_0\frac{\partial\psi}{\partial z} = \frac{2}{c}\frac{\partial^2\psi}{\partial z'\partial\tau} \tag{3}$$

where $\nabla^2_{x,y} = \frac{\partial^2}{\partial x^2} + \frac{\partial^2}{\partial y^2}$ and $k_0 = \frac{\omega_0}{c}$. In equation (3) we have considered the paraxial approximation and therefore dropped $\frac{\partial^2}{\partial z^2}$ term. Equation (3) governs the propagation of the pulsed paraxial optical beams in free space. Introducing a dimensionless variable $\tau' = \tau/T'$, where $T'$ is input pulse duration, we can write equation (3) as

$$\nabla^2_{x,y}\psi - 2ik_0\frac{\partial}{\partial z}\left[\psi + \frac{1}{2\pi i}\left(\frac{T_0}{T'}\right)\frac{\partial\psi}{\partial\tau'}\right] = 0. \tag{4}$$

In above equation, if pulse duration $T'$ is much longer than oscillation period $T_0$, the term $\frac{T_0}{T'}$ would be very small and then second term in equation (4) can be neglected. In this condition only the factorized solutions for the above equation exist. However, if optical pulse contains only a few optical cycles, the second term in equation (4) cannot be neglected and we witness spatio-temporal coupling. A family of non-separable solutions of equation (4) in this condition exists. As a solution to eqn. (4), space-time coupled nearly gaussian pulse has already been reported in literature. However, this pulse also contains $exp\ (r^4)$ term in transverse direction, which makes the pulse in the radial direction unbounded and nonintegrable [10]. On other hand, Poisson spectrum pulse also known as power exponential (PE) pulse constitutes one of the bounded solutions to the above $(3 + 1)$ dimensional paraxial wave equation [11,19]. The mathematical expression of the PE spectrum is given as,

$$\hat{a}_\omega = \frac{\pi}{\Gamma(\alpha+1/2)}\frac{\alpha^{\alpha+1/2}}{\bar{\omega}}\left(\frac{\omega}{\bar{\omega}}\right)^{\alpha-1/2} e^{\frac{-\omega}{\bar{\omega}}\alpha} e^{-i\varphi}, \tag{5}$$

where $\bar{\omega}$ is the mean frequency of the pulse, $\varphi$ is arbitrary phase and the parameter $\alpha > 1/2$. In the PE pulse, mean frequency $\bar{\omega} = \frac{\int_0^\infty |\hat{a_\omega}|^2 \omega d\omega}{\int_0^\infty |\hat{a_\omega}|^2 d\omega}$ appears explicitly. The temporal shape of the PE pulse with PE spectrum is expressed as,

$$a(t) = \frac{1}{\pi}\int_0^\infty \hat{a}_\omega e^{-i\omega t} d\omega = \left(\frac{-i\alpha}{\bar{\omega}t - i\alpha}\right)^{\alpha+1/2} e^{-i\varphi}. \tag{6}$$

The shape of the PE pulse is completely determined by $\alpha$ and scaled by $\bar{\omega}$. Upon increasing the value of $\alpha$, equation (6) resembles to Gaussian enveloped pulse with duration $\Delta t = \sqrt{2\alpha}/\bar{\omega}$ and $\varphi$ becomes the carrier envelope phase.

To experimentally generate ultrashort optical vortices, some technical challenges need to be overcome such as spatial, group velocity, and topological charge dispersions [30–32]. Suppose we synthesized the ultrashort optical vortices by addressing all these technical issues. The analytical complex representation of thus synthesized pulsed vortex can be given as

$$E(r,\phi,z,\tau) = \frac{1}{\pi}\int_0^\infty \hat{E}_\omega(r,\phi,z) e^{-i\omega\tau} d\omega, \tag{7}$$

where $\hat{E}_\omega(r,\phi,z)$ are LG beams of different frequencies and same topological charge *l*, which can mathematically be expressed as,

$$\hat{E}_\omega(r,\phi,z) = \hat{a}_\omega \frac{e^{-i(|l|+1)\Psi_\omega(z)} e^{-il\phi}}{\sqrt{1+\left(\frac{z}{z_{R,\omega}}\right)^2}}\left(\frac{\sqrt{2}r}{s_\omega(z)}\right)^{|l|} e^{i\omega r^2/[2cq_\omega(z)]}. \tag{8}$$

The beams possess identical topological charges *l* and zero radial orders. In above equation, we have omitted the LG polynomial $\left(L_p^{|l|}\left(\frac{2r^2}{s_\omega^2(z)}\right)\right)$ as for $p = 0$, the polynomial value will always be unity. The parameter $\hat{a}_\omega$ is the spectral weight of LG beams. In above equations, $\tau = t - \frac{z}{c}$ is the local time, $\Psi_\omega(z) = tan^{-1}\left(\frac{z}{z_{R,\omega}}\right)$, $q_\omega(z) = z - iz_{R,\omega}$ is the complex beam parameter, $s_\omega(z) = s_\omega\sqrt{1+\left(\frac{z}{z_{R,\omega}}\right)^2}$ is the beam width, $s_\omega = \sqrt{\frac{2z_{R,\omega}c}{\omega}}$ is the waist of fundamental Gaussian beam at $z = 0$ and $z_{R,\omega}$ is the Rayleigh range. The complex beam parameter $q_\omega(z)$ is generally written as

$$\frac{1}{q_\omega(z)} = \frac{1}{R_\omega(z)} + \frac{i2c}{\omega s_\omega^2(z)}, \tag{9}$$

where $\frac{1}{R_\omega(z)} = \frac{z}{z^2+z_{R,\omega}^2}$ represents the curvature of the phase fronts. Now integrating equation (7) with equation (5) and equation (8) at $z = 0$ gives

$$E(r,\phi,0,\tau) = \left(\frac{\sqrt{2}r}{s_{\bar{\omega}}}\right)^{|l|}\left[\frac{-i\left(\alpha+\frac{|l|}{2}\right)}{\bar{\omega}\left(\tau - i\frac{r^2}{\bar{\omega}s_{\bar{\omega}}^2}\right)-i\alpha}\right]^{\alpha+\frac{|l|}{2}+\frac{1}{2}} e^{-i\varphi}, \tag{10}$$

where $s_{\bar{\omega}}$ is waist of Gaussian beam at $z = 0$.

The dynamics of ultrashort optical pulses up to single cycle regime is described by nonlinear envelope equation (NEE). This equation is derived by resorting to the approximation known as slowly evolving wave approximation (SEWA). In this approximation not only envelope of the pulse remains unchanged during the propagation as assumed in slowly varying envelope approximation (SVEA) but also carrier envelope phase remains unchanged during the propagation. The NEE is given by [29,33-34],

$$\frac{\partial A}{\partial z} = \frac{i}{2k}\left(1 + \frac{i}{\omega}\frac{\partial}{\partial\tau}\right)^{-1}\nabla_\perp^2 A - \frac{i\beta_2}{2}\frac{\partial^2 A}{\partial\tau^2} + i\frac{n_2 n_o \omega}{2\pi}\left(1 + \frac{i}{\omega}\frac{\partial}{\partial\tau}\right)|A|^2 A, \tag{11}$$

Where $A(r,t)$ is the envelope of the optical pulse, $\beta_2$ is the group velocity dispersion (GVD) and $\tau = t - \frac{z}{c}$ is the local time. The presence of operator $\left(1 + \frac{i}{\omega}\frac{\partial}{\partial \tau}\right)^{-1}$ in Laplacian term of equation (11) gives space-time focusing while its presence with Kerr nonlinear term leads to self-steeping. Owing to the cylindrical symmetry of the LG vortex, the Laplacian in equation (11) can be expressed as $\left(\frac{\partial^2}{\partial r^2} + \frac{1}{r}\frac{\partial}{\partial r} + \frac{1}{r^2}\frac{\partial^2}{\partial \phi^2}\right)$. To numerically solve equation (11), we have made it dimensionless by implementing the following transformations,

$$u = \frac{A}{A_0}, \rho = \frac{r}{s_0}, T = \frac{\tau}{\tau_0}, Z = \frac{z}{L_{DF}}. \quad (12)$$

Using above transformations, where $A_0$ is the amplitude normalization parameter, $s_0$ is the beam waist, and $\tau_0$ is the temporal width of the pulse. we get following dimensionless form of NEE,

$$\frac{\partial u}{\partial Z} = \frac{i}{4}\left(1 + \frac{i}{s}\frac{\partial}{\partial T}\right)^{-1} \nabla_\perp^2 u - i\frac{L_{DF}}{L_{DS}}\frac{\partial^2 u}{\partial T^2} + i\frac{L_{DF}}{L_{NL}}\left(1 + \frac{i}{s}\frac{\partial}{\partial T}\right)|u|^2 u, \quad (13)$$

where $L_{DF} = \frac{k s_0^2}{2}$ is diffraction length, $L_{DS} = \frac{\tau_0^2}{|\beta_2|}$ is dispersion length , $L_{NL} = \left(\frac{n_0 n_2 \omega |A_0|^2}{2\pi}\right)^{-1}$ is nonlinear length and $s = \omega\tau_0$ represents self-steeping parameter [33]. The three characteristic lengths $L_{DF}$, $L_{DS}$ and $L_{NL}$ represent respective strengths of diffraction, dispersion, and nonlinear interactions of the LG vortex within the medium. The ratio of $\frac{L_{DF}}{L_{NL}}$ represents the relative strength of Kerr nonlinearity. We have considered the propagation in fused silica and used following simulation parameters: $\frac{L_{DF}}{L_{DS}} = 0.07$, $s = 40$, $\beta_2 = 385 fs^2/cm$ and $n_2 = 2 \times 10^{-16} cm^2/W$ [29]. The *fwhm* of the input LG vortex pulse centred at $795 nm$ is considered to be 30 $fs$.

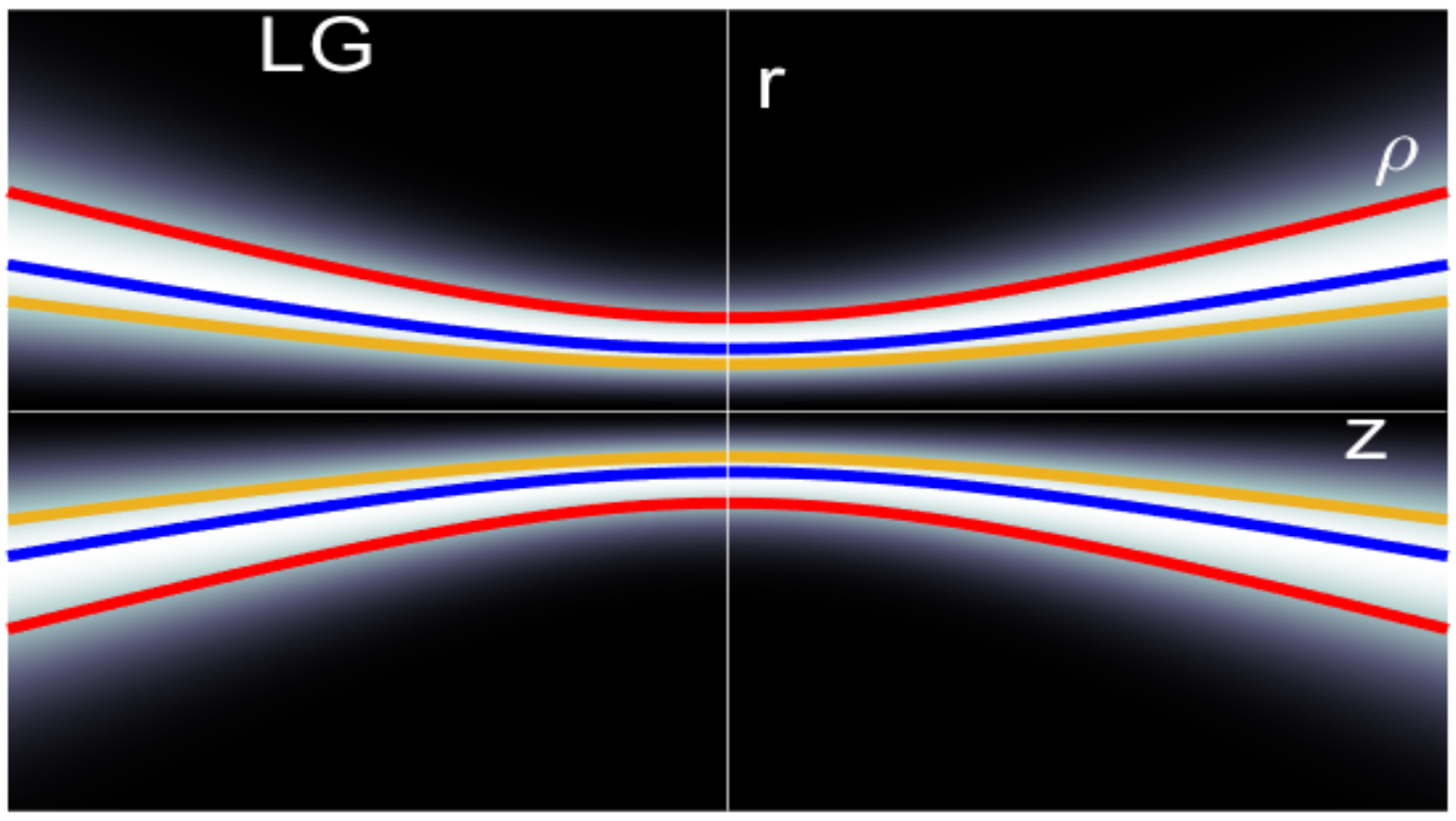

Fig. 1. Cross section of a LG beam. Blue ($\rho = \rho_p$), red ($\rho > \rho_p$), and yellow ($\rho < \rho_p$) curves represent revolution hyperboloids about the $z$ axis, called here caustic surfaces of LG beam. The blue curve ($\rho_p$) represents the bright caustic.

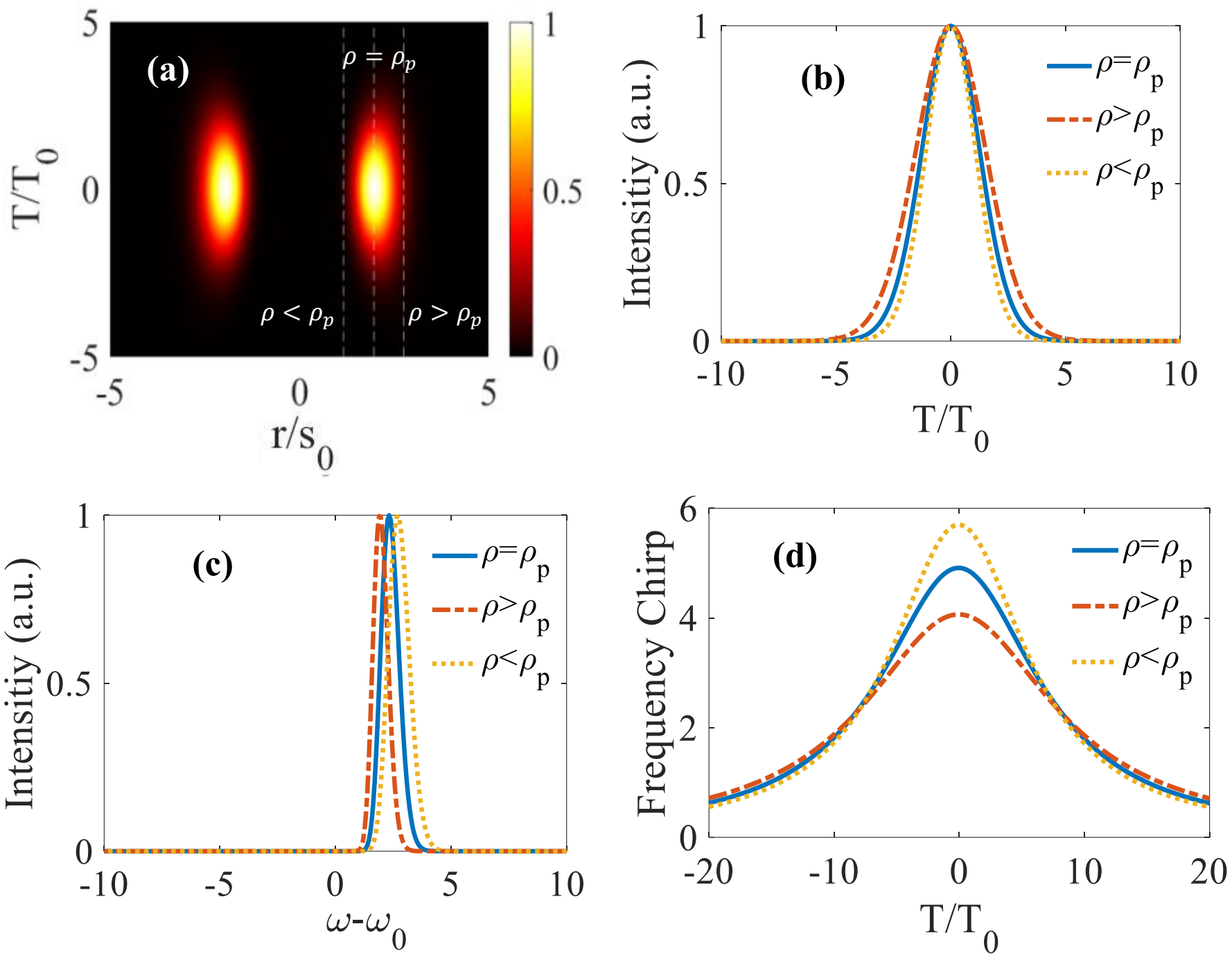

Fig. 2. Normalized input ultrashort LG vortex for $l = 8$. (a) Spatio-temporal profile, (b) normalized temporal profile at different caustics, (c) normalized spectral profile at different caustics, and (d) frequencies chirp at different caustics. Blue, red, and yellow caustics are at $\rho = 1.96\,unit$, $\rho = 2.76\,unit$ and $\rho = 1.16\,unit$ respectively.

Fig. 1 depicts the fluence distribution of the LG beam and the colorful curves on it represent the revolution hyperboloids or caustic surfaces whose revolution axis is $z$-axis. Fig. 2 shows the spatio-temporal transverse profile of the PE pulse and its temporal (Fig. 2 (b)), spectral (Fig. 2 (c)) and frequency chirp (Fig. 2 (d)) profiles at various caustics. In Fig. 2 (b), temporal broadening is observed as one moves radially outward from the phase singularity of the pulse. In spectral profile as shown in Fig. 2 (c), we observe redshift and blueshift in the spectrum of the pulse as one moves respectively radially outward and towards the phase singularity from the bright caustic of the pulse. Fig. 2(d) reveals the frequency chirp variation across various spatial regions of the pulse. Fig. 2(d) confirms the observations of Fig. 2(c).

## 3. Results and Discussion

### *3.1 Spatio-temporal* Evolution

In this section, we explore the spatio-temporal dynamics of ultrashort LG vortices with PE spectrum at various propagation distance in dispersive and nonlinear medium for $\frac{L_{DF}}{L_{NL}} = 1$. In Fig. 3(a), we observe that spatio-temporal profile of the pulse becomes tilted (as compared to input as shown in Fig. 2 (a)) at propagation distance $\frac{z}{L_{DF}} = 0.5$ and the tilt of the pulse increases with increase in the propagation distance (see Figs. 3(b) and (c)). The observed tilt in the pulse comes from the spatially varying spectral distribution. As depicted in Fig. 2(c), a redshift in spectrum is observed as we move radially outward from the bright caustic of the pulse and the redshifted components diffract strongly as compared to the blueshifted components. The spatially dependent spectrum and space-time non-separability are responsible for the tilt in the spatio-temporal plane. We see that upon increasing the propagation distance, the pulse broadens both in space and time because at higher distances diffraction and dispersion comes into the play and peak intensity both in space and time decrease. In Figs. 3 (a-c), the three dashed vertical lines in each figure represent the three caustics. In the second row of Fig. (3), we witness the temporal evolution of the pulse across different caustics of the ultrashort LG vortex. Notably, as we move radially outward ($\rho > \rho_p$) from the bright caustic ($\rho = \rho_p$) or approach towards the singularity ($\rho < \rho_p$), an asymmetric profile is observed in the temporal domain.

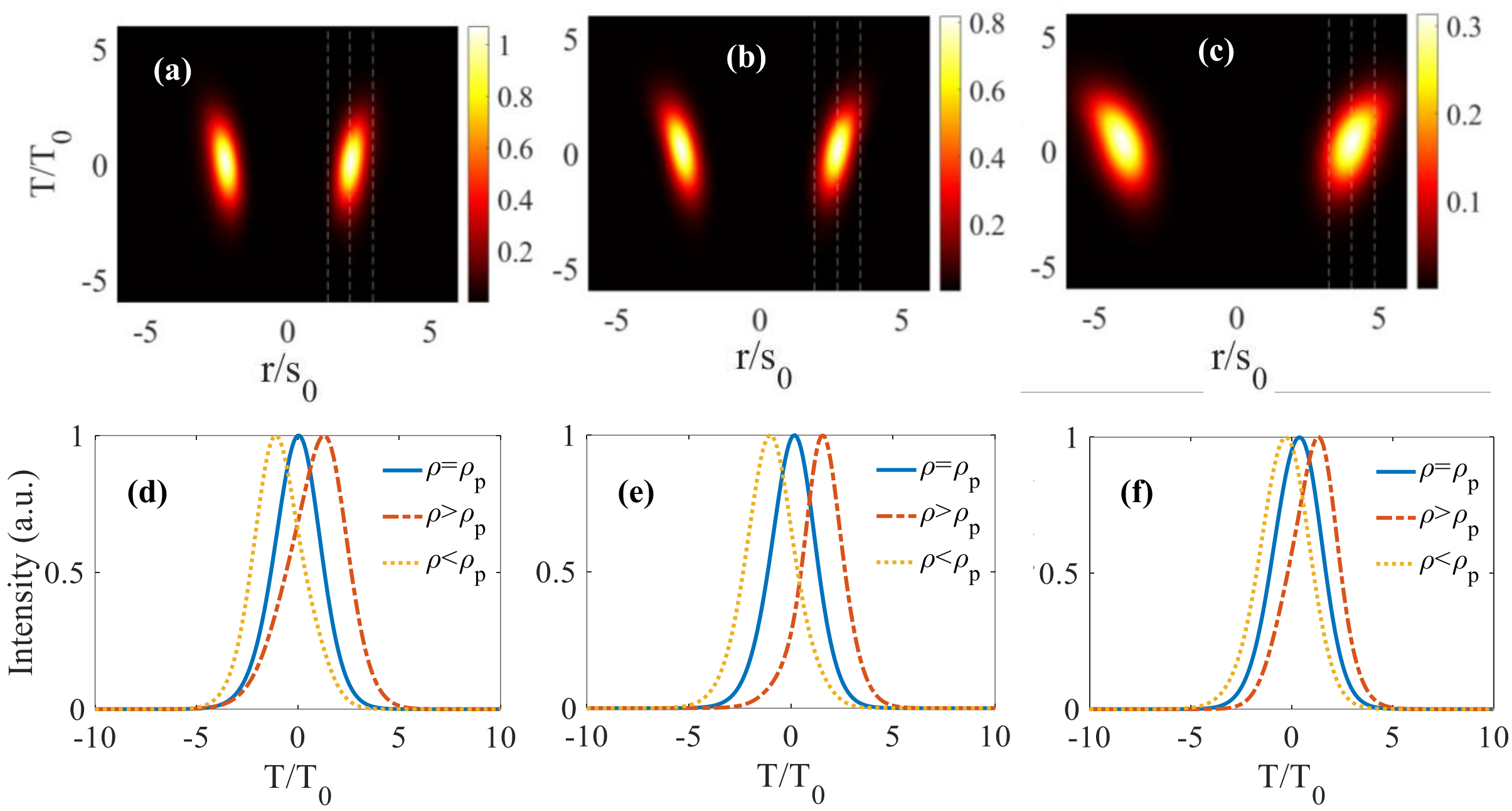

Fig. 3. Evolution of ultrashort LG vortex with PE spectrum for $l = 8$ for $\frac{L_{DF}}{L_{NL}} = 1$. The first row represents the spatio-temporal evolution of the pulse, and second row depicts the temporal evolution at various caustics of LG vortex. In the first row, the three dashed vertical lines in each Fig. represent three caustics. In all the Figs. of second row, we have normalized each graph with their respective maxima. Fig. (a) and Fig. (d) depict the evolution at $\frac{z}{L_{DF}} = 0.5$, Fig. (b) and Fig. (e) at $\frac{z}{L_{DF}} = 1$ and Fig. (c) and Fig. (f) at $\frac{z}{L_{DF}} = 2$. At $\frac{z}{L_{DF}} = 0.5$ blue, red, and yellow caustics are at $\rho = 2.2\, unit$, $\rho = 3.0\, unit$ and $\rho = 1.4\, unit$ respectively. At $\frac{z}{L_{DF}} = 1$ blue, red, and yellow caustics are at $\rho = 2.76\, unit$, $\rho = 3.56\, unit$ and $\rho = 1.96\, unit$ respectively. At $\frac{z}{L_{DF}} = 2$ blue, red, and yellow caustics are at $\rho = 4.04\, unit$, $\rho = 4.84\, unit$ and $\rho = 3.24\, unit$ respectively.

In Fig. (4), we have explored the dynamics of the pulse for higher nonlinear strength i.e., for $\frac{L_{DF}}{L_{NL}} = 5$. We observe in Fig. 4(a) that upon increasing the strength of nonlinearity the optical pulse gets compressed in space and time and that leads to splitting as observed in the second row of Fig. 4. The splitting in the pulse happens because $L_{DS} = 2.3$cm, $L_{DF} = 0.158$cm and $L_{NL} = 0.031$cm, therefore relatively stronger focusing Kerr nonlinearity pushes off-axis energy in time and space to flow toward center rapidly, as a result pulse gets compressed in both the domains and peak intensity of the pulse increases (compare the colorbars in Fig. 2(a) and Fig. 4(a). With spatial compression diffraction becomes stronger and it starts pushing the energy away from the center [35]. This leads to spatial splitting as depicted in Figs. 4(b) and (c). Similarly in temporal domain self-focusing causes off-axis energy to flow toward center rapidly, but GVD causes this energy to spread away from the center. This continues until self-focusing collapse terminates, which results in temporal pulse splitting [36]. As shown in Fig. 4(b), further propagation reveals a strong asymmetric splitting of the pulse, which is attributed to the previously discussed spatially dependent spectrum and space-time nonseparability. Upon further propagation (Fig. 4(c)) compression at the trailing edge of the optical pulse is seen. This occurs because of diffraction and Kerr nonlinearity working together on the high nonlinearity together with spatially dependent spectrum and space-time nonseparability of the vortex pulse. The second row of Fig. (4) depicts the temporal evolution at the different caustics of the LG vortices. Notably, we observe that upon moving radially away from the bright caustics, pulse compression and splitting dominate.

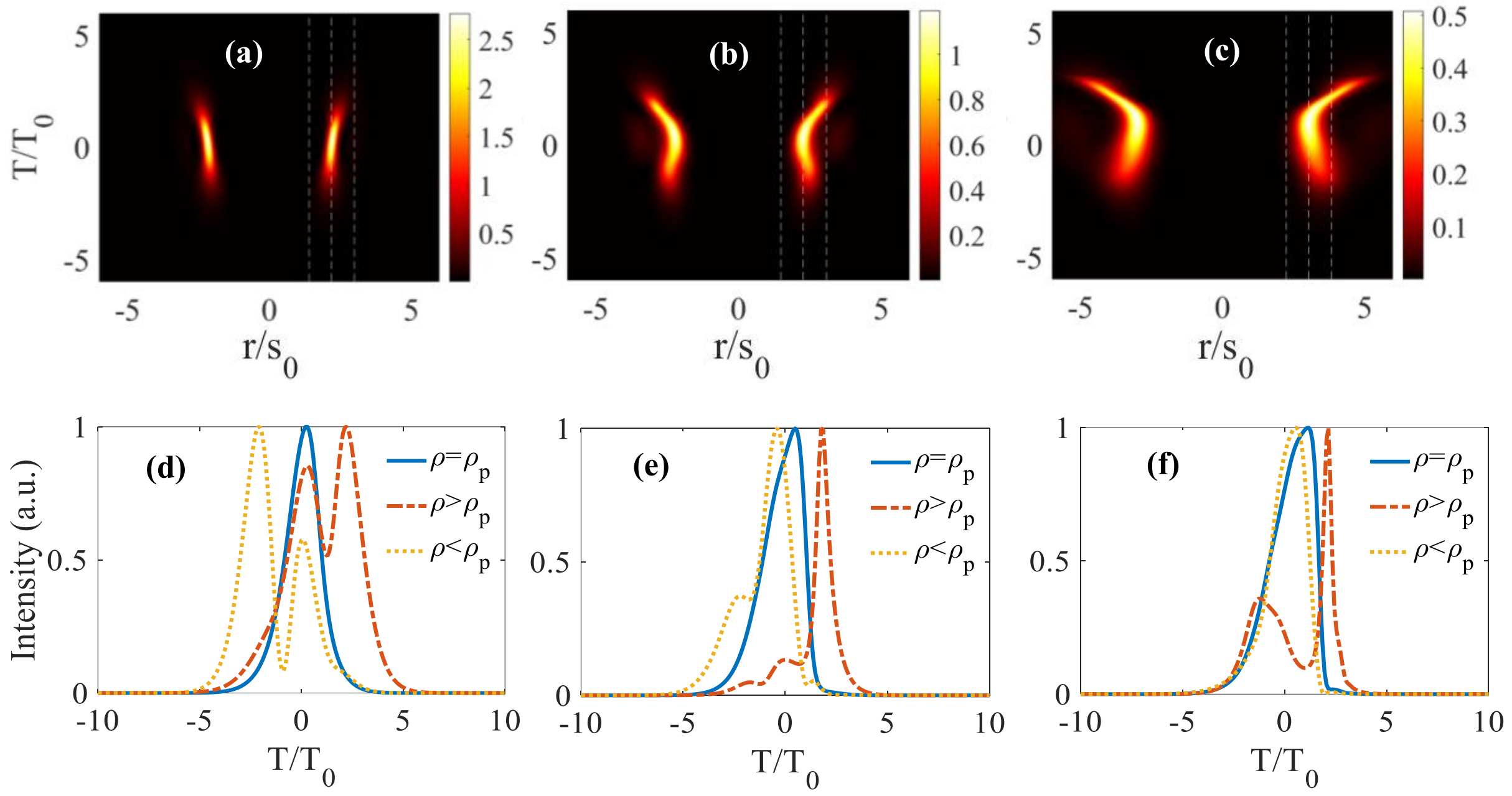

Fig. 4. Evolution of ultrashort LG vortex with PE spectrum for $l = 8$ with strength of nonlinearity $\frac{L_{DF}}{L_{NL}} = 5$. The first row represents the spatio-temporal evolution of the pulse, and the second row depicts the temporal evolution at various caustics of LG vortices. In the first row, the three dashed vertical lines in each Figs. represent three caustics. In all the Figs. of second row, we have normalized each graph with their respective maxima. Fig (a) and Fig. (d) depict the evolution at $\frac{z}{L_{DF}} = 0.5$, Fig. (b) and Fig. (e) at $\frac{z}{L_{DF}} = 1$ and Fig. (c) and Fig. (f) at $\frac{z}{L_{DF}} = 2$. At $\frac{z}{L_{DF}} = 0.5$ blue, red, and yellow caustics are at $\rho = 2.20\ unit$, $\rho = 3.00\ unit$ and $\rho = 1.40\ unit$ respectively. At $\frac{z}{L_{DF}} = 1$ blue, red, and yellow caustics are at $\rho = 2.28\ unit$, $\rho = 3.08\ unit$ and $\rho = 1.48\ unit$ respectively. At $\frac{z}{L_{DF}} = 2$ blue, red, and yellow caustics are at $\rho = 3.04\ unit$, $\rho = 3.84\ unit$ and $\rho = 2.24\ unit$ respectively.

### *3.2 Spectral Evolution*

In this section, we explore the spectral evolution of ultrashort LG vortex at its different caustics. As shown in Fig. 5(a), for $\frac{L_{DF}}{L_{NL}} = 1$ and $z = 0.5L_{DF}$ as we move radially outward from the bright caustic, the ultrashort LG vortex spectrum exhibits redshift, while on moving towards the phase singularity a blueshift in spectrum is observed. Upon further propagation at $z = L_{DF}$ (Fig. 5(b)), the spectra of the different spatial regions in the ultrashort LG vortex coincides with that of the bright caustic. However, upon further propagation at $z = 2L_{DF}$, we observe spectral characteristics opposite to what is seen in Fig. 5(a). Upon increasing the strength of the nonlinearity to $\frac{L_{DF}}{L_{NL}} = 5$, (second row of Fig. 5), we observe significant variation in the spectrum of the ultrashort LG vortices. We observe in Fig. 5(e) that at the bright caustic ($\rho = \rho_p$) of the LG vortices, the spectrum has become asymmetrical. Approaching towards the singularity ($\rho < \rho_p$) amplifies this asymmetry. Further, moving radially outward from the bright caustic ($\rho > \rho_p$) gives minimal spectral changes. As propagation continues, nonlinear effects become increasingly pronounced, leading to complex spectral changes across different spatial regions of the LG vortex.

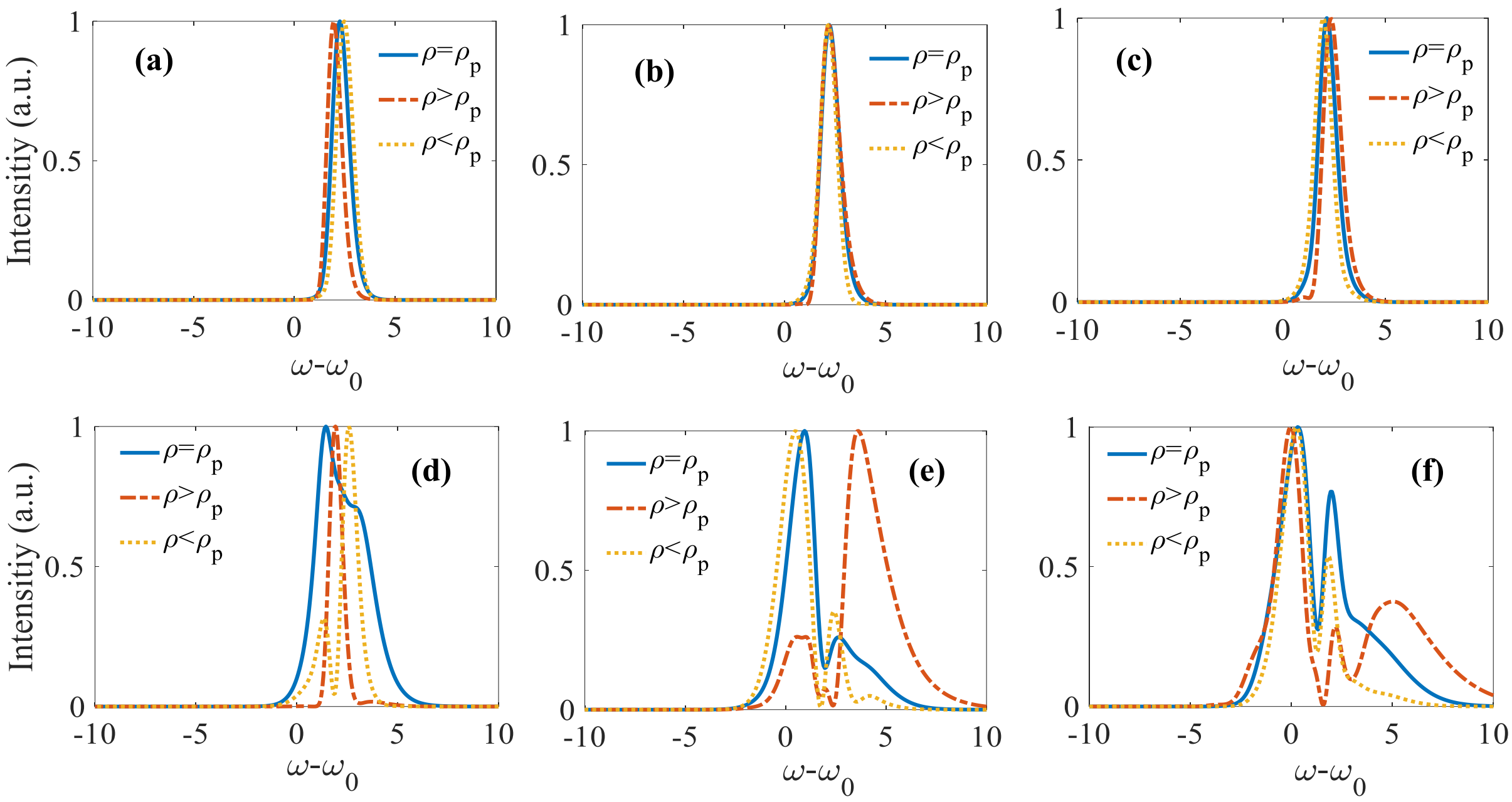

Fig. 5. Spectral evolution of ultrashort LG vortex with PE spectrum across various spatial region for $l = 8$. In first column $\frac{z}{L_{DF}} = 0.5$, second column $\frac{z}{L_{DF}} = 1$ and in third column $\frac{z}{L_{DF}} = 2$. In first row $\frac{L_{DF}}{L_{NL}} = 1$ and in second row $\frac{L_{DF}}{L_{NL}} = 5$. Blue curve ($\rho = \rho_p$) represents the bright caustic, red ($\rho > \rho_p$) represents away from the bright caustic and ($\rho < \rho_p$) represents caustic towards the singularity. In all the Figs., we have normalized each graph with their respective maxima. In first row at $\frac{z}{L_{DF}} = 0.5$ blue, red, and yellow caustics are at $\rho = 2.2\,unit$, $\rho = 3.0\,unit$ and $\rho = 1.4\,unit$ respectively. At $\frac{z}{L_{DF}} = 1$ blue, red, and yellow caustics are at $\rho = 2.76\,unit$, $\rho = 3.56\,unit$ and $\rho = 1.96\,unit$ respectively. At $\frac{z}{L_{DF}} = 2$ blue, red, and yellow caustics are at $\rho = 4.04\,unit$, $\rho = 4.84\,unit$ and $\rho = 3.24\,unit$ respectively. In second row, at $\frac{z}{L_{DF}} = 0.5$ blue, red, and yellow caustics are at $\rho = 2.20\,unit$, $\rho = 3.00\,unit$ and $\rho = 1.40\,unit$ respectively. At $\frac{z}{L_{DF}} = 1$ blue, red, and yellow caustics are at $\rho = 2.28\,unit$, $\rho = 3.08\,unit$ and $\rho = 1.48\,unit$ respectively. At $\frac{z}{L_{DF}} = 2$ blue, red, and yellow caustics are at $\rho = 3.04\,unit$, $\rho = 3.84\,unit$ and $\rho = 2.24\,unit$ respectively.

### *3.3 Frequency Chirp Evolution*

In this section, we study the evolution dynamics of frequency chirp [37,38] of ultrashort LG vortex at different spatial regions across the pulse. The chirp of the optical pulse measures the time dependence of its instantaneous frequency. For Gaussian pulses, its nature is linear in dispersive medium while nonlinearity imposes linear chirp only over a large central region of the pulse. The frequency chirp of the PE pulse peaks at the pulse center, i.e. blueshifted spectrum of the pulse lies in the central region of the pulse while leading and trialing edge of the pulse has redshifted components. Fig. 6(a) illustrates the frequency chirp of the PE pulse at $\frac{\mathrm{z}}{\mathrm{L_{DF}}} = 0.5$. Notably, we observe nominal variations in the shape of the frequency chirp while examining it at three different transverse spatial location-at the bright caustic, radially outward from it, and toward the phase singularity. However, as we propagate further i.e. at $\frac{z}{L_{DF}} = 1$, the frequency chirp changes slightly for all three spatial regions. At the bright caustic (blue curve), a second peak in the frequency chirp emerges, a feature also observed for spatial region (yellow dashed curve) close to the phase singularity. As the pulse furthers, we observe typical signatures of the self-phase modulation on frequency chirp.

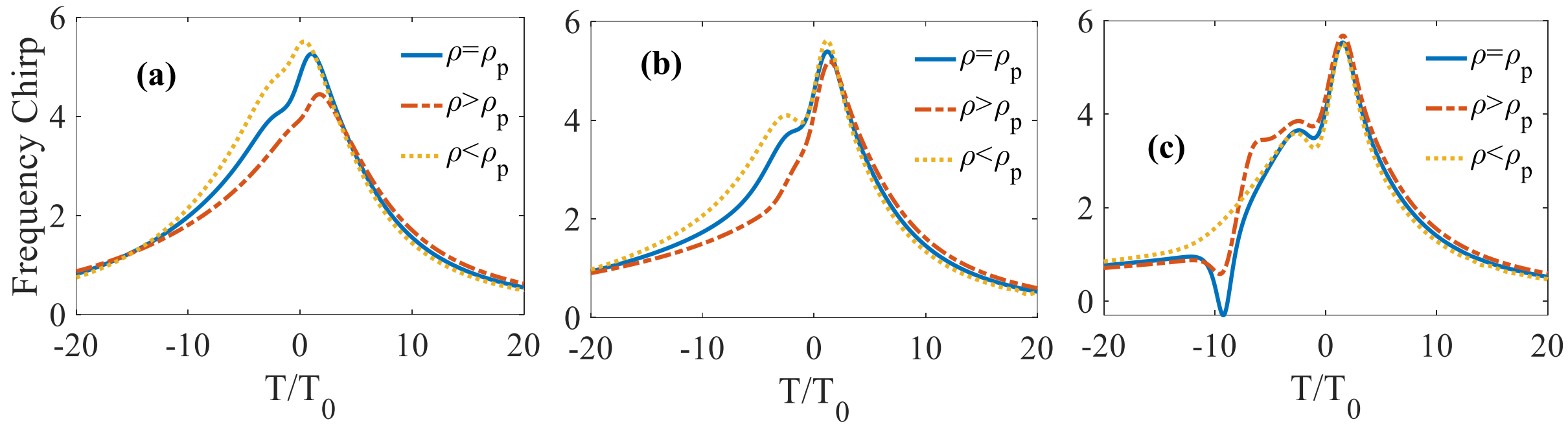

Fig. 6. Frequency chirp evolution of ultrashort LG vortex with P. E. spectrum across various spatial region for $l = 8$ at different propagation distances for ( $\frac{L_{DF}}{L_{NL}} = 1$). (a) $\frac{z}{L_{DF}} = 0.5$, (b) $\frac{z}{L_{DF}} = 1$ and (c) $\frac{z}{L_{DF}} = 2$. At $\frac{z}{L_{DF}} = 0.5$ blue, red, and yellow caustics are at $\rho = 2.2\ unit$, $\rho = 3.0\ unit$ and $\rho = 1.4\ unit$ respectively. At $\frac{z}{L_{DF}} = 1$ blue, red, and yellow caustics are at $\rho = 2.76\ unit$, $\rho = 3.56\ unit$ and $\rho = 1.96\ unit$ respectively. At $\frac{z}{L_{DF}} = 2$ blue, red, and yellow caustics are at $\rho = 4.04\ unit$, $\rho = 4.84\ unit$ and $\rho = 3.24\ unit$ respectively.

Upon increasing the strength of nonlinearity, we observe frequency chirp similar to that introduced by self-phase modulation as shown in Fig. 7. This is due to the dominance of SPM imposed chirp over pulse intrinsic chirp. This ultimately results into a pulse with positive chirp in pulse central region as shown in Fig. 7(c).

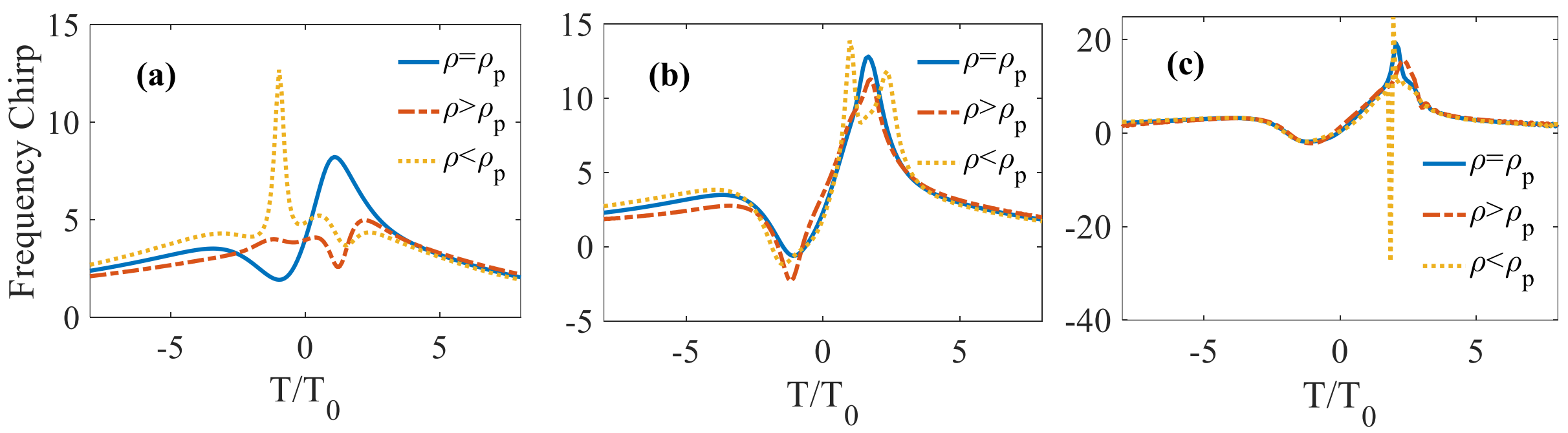

Fig. 7. Frequency chirp evolution of ultrashort LG vortex with PE spectrum across various spatial region for $l = 8$ at different propagation distances for ( $\frac{L_{DF}}{L_{NL}} = 5$). At $\frac{z}{L_{DF}} = 0.5$ blue, red, and yellow caustics are at $\rho = 2.20\ unit$, $\rho = 3.00\ unit$ and $\rho = 1.40\ unit$ respectively. At $\frac{z}{L_{DF}} = 1$ blue, red, and yellow caustics are at $\rho = 2.28\ unit$, $\rho = 3.08\ unit$ and $\rho = 1.48\ unit$ respectively. At $\frac{z}{L_{DF}} = 2$ blue, red, and yellow caustics are at $\rho = 3.04\ unit$, $\rho = 3.84\ unit$ and $\rho = 2.24\ unit$ respectively.

## 4. Conclusion

In this paper, we have studied the dynamics of space-time non-separable ultrashort LG vortex with PE spectrum in dispersive and nonlinear media at different transverse spatial regions of ultrashort vortex by employing NEE. The spatio-temporal evolution of the ultrashort vortex pulse showed asymmetry in spatio-temporal plane due to spatially dependent spectrum. Furthermore, for stronger nonlinearity and at larger propagation distances, we observed compression and splitting in the pulse in spatio-temporal plane induced by diffraction and Kerr nonlinearity. Different spatial regions of the ultrashort LG vortex witnessed different kinds of temporal asymmetric splitting. The Spectral evolution of the pulse revealed a redshifted spectrum moving radially outward from the bright caustic and a blueshifted spectrum was observed on moving towards the phase singularity at lower nonlinearity levels and shorter distances. As propagation distance increases, the spectrum undergoes further changes, with a noticeable redshift towards the phase singularity and a blueshift radially outward from the bright caustic. Upon increasing the strength of nonlinearity and propagation distance, we observed complex spectral evolution at different spatial regions. The frequency chirp of ultrashort LG vortices with PE spectrum has also been investigated. We found that for weaker nonlinearity, the shape of the frequency chirp remains unaffected however upon increasing the strength of the nonlinearity frequency chirp resembles to that introduced by self-phase modulation.

This study may hold various applications in optical communication, compressing femtosecond LG vortices, generating supercontinuum of ultrashort LG vortices, improving temporal resolution in ultrafast spectroscopy, and OAM based tunable optical tweezers.

**Acknowledgements**

Shakti Singh wishes to acknowledge the University Grant Commission (India) for the financial support. This work is supported by the research grant CRG/2022/007736 from SERB (India).

**Disclosure.** The authors declare no conflicts of interest.

**Data availability.** Data underlying the results presented in this paper are not publicly available at this time but may be obtained from the authors upon reasonable request.

**References:**

1. A. M. Shaltout, K. G. Lagoudakis, J. van de Groep, S. J. Kim, J. Vučković, V. M. Shalaev, and M. L. Brongersma, Spatiotemporal light control with frequency-gradient metasurfaces, Science **365**, 374 (2019).

2. Z. Qiao, Z. Wan, G. Xie, J. Wang, L. Qian, and D. Fan, Multi-vortex laser enabling spatial and temporal encoding, PhotoniX **1**, 13 (2020).

3. Z. Nie, C. H. Pai, J. Hua, C. Zhang, Y. Wu, Y. Wan, F. Li, J. Zhang, Z. Cheng, Q. Su, S. Liu, Y. Ma, X. Ning, Y. He, W. Lu, H. H. Chu, J. Wang, W. B. Mori, and C. Joshi, Relativistic single-cycle tunable infrared pulses generated from a tailored plasma density structure, Nat. Photonics **12**, 489 (2018).

4. M. Malinauskas, A. Žukauskas, S. Hasegawa, Y. Hayasaki, V. Mizeikis, R. Buividas, and S. Juodkazis, "Ultrafast laser processing of materials: From science to industry," Light Sci. Appl. **5**, e16133 (2016).

5. Y. Shen, X. Wang, Z. Xie, C. Min, X. Fu, Q. Liu, M. Gong, and X. Yuan, Optical vortices 30 years on: OAM manipulation from topological charge to multiple singularities, Light Sci Appl **8**, (2019).

6. Y. Shen, A. Zdagkas, N. Papasimakis, and N. I. Zheludev, Measures of space-time nonseparability of electromagnetic pulses, Phys. Rev. Res. **3**, 013236 (2021).

7. S. Akturk, X. Gu, P. Bowlan, and R. Trebino, Spatio-temporal couplings in ultrashort laser pulses, J. Opt. **12**, 093001 (2010).

8. M. A. Porras, Diffraction effects in few-cycle optical pulses, Phys. Rev. E **65**, 026606 (2002).

9. E. M. Belenov and A. V. Nazarkin, Transient diffraction and precursorlike effects in vacuum, J. Opt. Soc. Am. A **11**, 168 (1994).

10. Z. Wang, Z. Zhang, Z. Xu, and Q. Lin, Space-time profiles of an ultrashort pulsed Gaussian beam, IEEE J Quantum Electron **33**, 566 (1997).

11. M. A. Porras, "Nonsinusoidal few-cycle pulsed light beams in free space," J. Opt. Soc. Am. B **16**, 1468 (1999).

12. L. Allen, M. W. Beijersbergen, R. J. C. Spreeuw, and J. P. Woerdman, Orbital angular momentum of light and the transformation of Laguerre-Gaussian laser modes, Phys. Rev. E **45**, 31 (2016).

13. S. Singh, S. K. Mishra, and A. K. Mishra, Ring Pearcey vortex beam dynamics through atmospheric turbulence, J. Opt. Soc. Am. B **40**, 2287 (2023).

14. S. Singh and A. K. Mishra, Spatio-temporal evolution dynamics of ultrashort Laguerre-Gauss vortices in a dispersive and nonlinear medium, J. Opt. **24**, 075501 (2022).

15. J. Wang, J. Y. Yang, I. M. Fazal, N. Ahmed, Y. Yan, H. Huang, Y. Ren, Y. Yue, S. Dolinar, M. Tur, and A. E. Willner, Terabit free-space data transmission employing orbital angular momentum multiplexing, Nat Photonics **6**, 488 (2012).

16. D. G. Grier, A revolution in optical manipulation, Nature **424**, 810 (2003).

17. A. Jesacher, S. Fürhapter, S. Bernet, and M. Ritsch-Marte, Shadow effects in spiral phase contrast microscopy, Phys. Rev. Let.t **94**, 233902 (2005).

18. G. Molina-Terriza, J. P. Torres, and L. Torner, Management of the Angular Momentum of Light: Preparation of Photons in Multidimensional Vector States of Angular Momentum, Phys. Rev. Lett. **88**, 013601 (2002).

19. M. A. Porras, Effects of the Coupling between the Orbital Angular Momentum, and the Temporal Degrees of Freedom in the Most Intense Ring of Ultrafast Vortices, Appl. Sci. **10**, 1957 (2020).

20. M. A. Porras, Upper Bound to the Orbital Angular Momentum Carried by an Ultrashort Pulse, Phys. Rev. Lett. **122**, 123904 (2019).

21. Y. Chen, X. Feng and C. Liu, Generation of Nonlinear Vortex Precursors, Phys. Rev. Lett. 117, 023901(2016)

22. M. A. Porras, C. Conti, and C. Conti, Couplings between the temporal and orbital angular momentum degrees of freedom in ultrafast optical vortices, Phys. Rev. A **101**, 63803 (2020).

23. M. Ornigotti, C. Conti, and A. Szameit, Effect of Orbital Angular Momentum on Nondiffracting Ultrashort Optical Pulses, Phys. Rev. Lett. **115**, 100401 (2015).

24. M. A. Porras, Ultrashort pulsed Gaussian light beams, Phys. Rev. E **58**, 1086 (1998).

25. M. A. Porras, Effects of orbital angular momentum on few-cycle and sub-cycle pulse shapes: coupling between the temporal and angular momentum degrees of freedom, Opt. Lett. **44**, 2538 (2019).

26. Y. Shen, C. R. Guzmán, Nonseparable States of Light: From Quantum to Classical, Laser Photonics Rev. 16, 2100533 (2022).

27. A. Zdagkas, C. McDonnell, J. Deng, Y. Shen, G. Li, T. Ellenbogen, N. Papasimakis, and N. I. Zheludev, Observation of toroidal pulses of light, Nat. Photonics, **16**, 523 (2022).

28. Y. Shen, N. Papasimakis, N. I. Zheludev, Nondiffracting supertoroidal pulses and optical “Kármán vortex streets”, Nat. Commun. **15**, 4863 (2024).

29. J. K. Ranka and A. L. Gaeta, Breakdown of the slowly varying envelope approximation in the self-focusing of ultrashort pulses, Opt. Lett. **23**, 244 (1998).

30. K. Bezuhanov, A. Dreischuh, G. G. Paulus, M. G. Schätzel, and H. Walther, Vortices in femtosecond laser fields, Opt. Lett. **29**, 1942 (2004).

31. I. Zeylikovich, H. I. Sztul, V. Kartazaev, T. Le, and R. R. Alfano, Ultrashort Laguerre-Gaussian pulses with angular and group velocity dispersion compensation, Opt. Lett. **32**, 2025 (2007)

32. Y. Tokizane, K. Oka, and R. Morita, Supercontinuum optical vortex pulse generation without spatial or topological-charge dispersion, Opt. Express **17**, 14517 (2009).

33. T. Brabec and F. Krausz, Nonlinear optical pulse propagation in the single-cycle regime, Phys. Rev. Lett. **78**, 3282 (1997).

34. L. Xu, D. Li, J. Chang, T. Xi, Z. Hao, Few-cycle vortex beam generated from self-compression of midinfrared femtosecond vortices in thin plates, Phys. Rev. A, **106**, 053516 (2022).

35. A. Kumar and A. K. Mishra, Anomalous self-steepening, temporal pulse splitting and ring formation in a left-handed metamaterial with cubic nonlinearity, J. Opt. Soc. Am. B, **29**, 1330-1337 (2012).

36. J. E. Rothenberg, Pulse splitting during self-focusing in normally dispersive media, Opt. Lett. **17**, 583-585 (1992)

37. G. P. Agrawal, Nonlinear Fiber Optics, Academic Press, Fifth Edition (2013).

38. R. Trebino, Frequency-Resolved Optical Gating: The Measurement of Ultrashort Laser Pulses